\documentclass[a4paper,aps,prl,showpacs,amssymb,twocolumn,superscriptaddress]{revtex4}
\usepackage{graphicx}

\begin{document}
\date{\today}

\title{Competition Among Companies: Coexistence and Extinction}

\author{Marcelo Kuperman}
\email{kuperman@cab.cnea.gov.ar} \affiliation{Centro At{\'o}mico
Bariloche and Instituto Balseiro, 8400 S. C. de Bariloche,
Argentina}

\author{Horacio S. Wio}
\email{ wio@imedea.uib.es,
wio@cab.cnea.gov.ar}\affiliation{Centro At{\'o}mico Bariloche and
Instituto Balseiro, 8400 S. C. de Bariloche, Argentina}
 \affiliation{Instituto Mediterr\'aneo de Estudios Avanzados, IMEDEA
(CSIC-UIB),Campus UIB, 07071-Palma de Mallorca, Spain}

\begin{abstract}
We study a spatially homogeneous model of a market where several
agents or companies compete for a wealth resource. In analogy with
ecological systems the simplest case of such models shows a kind
of ``competitive exclusion" principle. However, the inclusion of
terms corresponding for instance to ``company efficiency" or to
(ecological) ``intracompetition" shows that, if the associated
parameter overcome certain threshold values, the meaning of
``strong" and ``weak" companies should be redefined. Also, by
adequately adjusting such a parameter, a company can induce the
"extinction" of one or more of its competitors.
\end{abstract}

\pacs{89.65.Gh, 87.23.Cc, 05.45.-a}

\maketitle

\section{Introduction}

During the last few years we have witnessed a wealth of work on
the application of methods of statistical physics to the study of
economic problems configuring what some authors called {\it
econophysics} \cite{uno,dos,tres,cuatro}. Within this framework a
great deal of effort was dedicated to the analysis of economic
data ranging from stock exchange fluctuations \cite{uno},
production models, size distribution of companies, the appearance
of money, effects of control on the market to market critical
properties \cite{cinco}. Another problem that attracted enormous
interest was the origin of power (Paretto) laws, and lognormal
distribution with power law tails, for the income of individuals,
wealth distribution, debt of bankrupt companies \cite{SS}. An
interesting source of several mathematical descriptions and models
used in economical and sociological contexts can be found in
\cite{haag}.

Here, our interest is the study, in a deterministic way, of
aspects of the competition and coexistence of agents or companies
in a common market. We present a simple ``toy" model describing,
in analogy to some ecological problems, a situation of competition
among several companies. Within a ``Malthusian-like" model we
analyze the effect of a kind of {\it intracompetitive}
contribution on the possibility of companies coexistence in a
certain market, and the changing leadership role (measure through
some wealth parameter) between, according to the standard
ecological definition, ``strong" and ``weak" companies.

According to ecological studies, starting with Volterra's first
results on the mathematical theory of competition \cite{vol}, the
problem of competition and coexistence between species has been
analyzed and resumed within the {\it Competitive Exclusion
Principle} (or {\it Ecological Theorem}), that states: {\it $N$
species that compete for $n (<N)$ food resources, cannot coexist}
\cite{mur}. Several aspects of this problem have been analyzed by
different authors, emphasizing, for instance, the conflict between
the need to forage and the need to avoid competition; effects of
diffusion-mediated persistence. Generally, the system describing
competition between species can be represented by a set of $N$
differential equations for the species and $n$ for the resources.
As an example, with only one food resource we find only two stable
stationary solutions: the trivial one (extinction of all he
species), and that corresponding to the survival of only one
species, the ``strongest" one. There are also studies of the
problem related to the possibility of coexistence in the form of
wave--like solutions \cite{sch,wkvh}.

Here we adapt the model used in Refs.\cite{sch,wkvh} for the case
of a homogeneous competitive market with a unique wealth resource
and several firms. In the next Section we introduce the model and
some particular, instructive, solutions. Due to the difficulties
of finding analytical solutions for the general case, in Section
III we focus on a representative system with a small number (here
4) of firms, and analyze its behavior by numerical methods. In the
last Section we draw some conclusions.

\section{The Model}

We start describing the model we use, which is related to the one
used in Refs.\cite{sch,wkvh} for the study of coexistence in an
ecological framework in the form of wave--like solutions. Such a
model has been adapted to the problem of competition of $N$ agents
or companies for a unique common wealth resource. We indicate with
$n_j$ the ``size" or wealth-parameter representing the welfare of
the $j$-company ($j=1,...,N$) and with $M$ the total amount of
``wealth". The set of differential equations that we use to
describe the behavior (for the homogeneous case) of such a complex
system includes a ``Malthusian-like" birth-death equation
\cite{mal} for each company ($\beta _j\,M(t)\, n_j(t)\,\,$
corresponding to benefits coming from the wealth's share and $\,\,
-\alpha _j\, n_j(t)\,\,$ for the ``standard" losses--or costs-- of
the $j$-th company). We also include a contribution that
corresponds, in ecological language, to taking into account the
existence of some kind of {\it intracompetition}, that is {\bf if}
the $j$-th the company is alone and can get all the wealth $M$, it
can only grow up to a maximum bounded size, a behavior that can be
modelled by a Verhulst-like term \cite{mur}. This takes into
account the in ecological contexts so called {\it carrying
capacity} of the economic environment \cite{ccap,yad}. In economic
terms it could correspond to an increase of the company
efficiency, for instance, through the reduction of operative
internal costs, improved management, avoiding of competition
between different branches of the same firm, instrumentation of
new technologies, etc.

In addition, instead of assuming a constant finite wealth resource
as in the so called ``Total Wealth Conserved model" \cite{piav},
we consider that $M$ has its own dynamics. For the equation for
$M$, the economic wealth accessible to the companies that, in
order to simplify this initial analysis we assume is unique, we
also consider a ``Malthusian-like" behavior including: its
production (new resources and technologies, harvest and grain
production, etc) that we assume has a constant rate $Q$, and its
disappearance due to: (a) natural degradation or rotting of crops,
technologies becoming old, some resources being exhausted, that we
assume has a rate proportional to the total wealth amount $- G M$
(only a certain portion of $M$ disappears), (b) the share of each
company given by $- \beta _k\,n_k(t)\, M(t)$.

The set of equations is ($j=1,...,N$)
\begin{eqnarray} \label{sis0}
\frac{d}{dt} n_j(t)&=& [\beta _j\,M(t)-\alpha _j]\, n_j(t) -
\gamma _j \frac{n_j^2}{M} \nonumber\\
\frac{d}{dt}  M(t)&=&Q(t) - [G+ \sum \beta _k\,n_k(t)]\, M(t).
\end{eqnarray}

Similarly to what was discussed in \cite{wkvh} for the case of
only two species, we can here define a hierarchy from the
``strongest" (largest ratio between wealth share and standard
losses, i.e. largest $\frac{\beta _i}{\alpha _i}$) to the
``weakest" (smallest ratio) companies. Assuming the following
hierarchical order
\begin{equation}
\frac{\beta _1}{\alpha _1} >  \frac{\beta _2}{\alpha _2} > ... >
\frac{\beta _N}{\alpha _N},
\end{equation}
we have that $n _1$ is the strongest company while $n_N$ is the
weakest one. It is worth noting in passing that Eqs.(\ref{sis0})
resembles the form of multimode laser systems \cite{lasers},
making it possible to transfer some results from one system to the
other.

The stationary solutions results from taking $\frac{d}{dt} M =0$
and $\frac{d}{dt} n_j =0$ ($j=1,...,N$). We found
\begin{eqnarray} \label{stat1}
M_s &=& \frac{Q(t)}{G+ \sum \beta _k\,n_k^s} \\
0&=& \left( [\beta _j\,M_s-\alpha _j] - \gamma _j
\frac{n_j^s}{M_s} \right) \, n_j^s.
\end{eqnarray}
The last equation implies one of two possibilities
\begin{equation} \label{stat2}
n_j^s= 0, \,\,\,\,\,\,\, \mbox{or} \,\,\,\,\,\,\,
n_j^s=\frac{[\beta _j\,M_s-\alpha _j]\,M_s}{\gamma _j }.
\end{equation}

It is clear that for large $N$, to find the solution of this
system is not easy. In order to fix ideas we consider the
simplified case where, instead of the above indicated hierarchy,
we have that all companies are equivalent, that is
$$\beta _j = \beta, \,\,\,\, \alpha _j = \alpha, \,\,\,\, \gamma _j
= \gamma, $$ implying $$n_j^s= n_s\,\,\,\,\,\,\ \forall \,j.$$ In
this case we have
\begin{equation} \label{stat3}
n_s=\frac{[\beta\, Q - (\alpha /Q) (G + N \beta n_s)]\,
Q^2}{\gamma (G + N \beta n_s)^2},
\end{equation}
that can be rewritten as
\begin{equation} \label{stat4}
- \frac{\gamma N^2 \beta}{Q^2} n_s^3 - \frac{2\gamma G N
\beta}{Q^2} n_s^2 - \left( \frac{\alpha \beta N}{Q} + \frac{\gamma
G^2}{Q^2} \right) n_s + ( \beta - \frac{\alpha G}{Q}) = 0.
\end{equation}
It is possible to find under which conditions at least one
solution of Eq. (\ref{stat4}) is real. However, it is more
instructive to look for the behavior at small $n_s$ ($n_s \sim 0$)
as all the coefficients of $n_s^\nu$ with $\nu > 0$ are negative.
Hence, we can easily obtain that a solution, given by
$$n_s \approx \frac{\beta Q^2- \alpha G Q}{\alpha \beta N Q + \gamma
G^2},$$ exists (is positive) if $\beta Q > \alpha G.$ In this
case, as one of the associated eigenvalues is zero, a linear
stability analysis does not give a clear information about the
stability of the solution and it is necessary to resort to a more
refined analysis.

Another instructive case is to consider
$$\frac{\beta _1}{\alpha _1} >  \frac{\beta _2}{\alpha _2} = ... =
\frac{\beta _N}{\alpha _N}.$$ Here we reduce to essentially the
same situation studied in \cite{sch,wkvh}. In particular, it is
coincident with the situation studied in \cite{KW}, but now having
an ``effective" weak species given by $(N-1) n_j$, $j=2,N$. As in
\cite{KW}, and as discussed in detail latter for the case of
several firms, it is possible to find a coexistence region when
$\gamma _1$ overcomes some threshold value. In this case a linear
stability analysis shows a change in the stability of these
solutions.

As a general analytical study of our system even for $N$ not too
large is far beyond our interest, in the next section we focus on
a numerical approach for a case with a small, however
representative, value of $N$ analyzing some relevant situations.

\begin{figure}[h]
\centering \resizebox{\columnwidth}{!}{\includegraphics{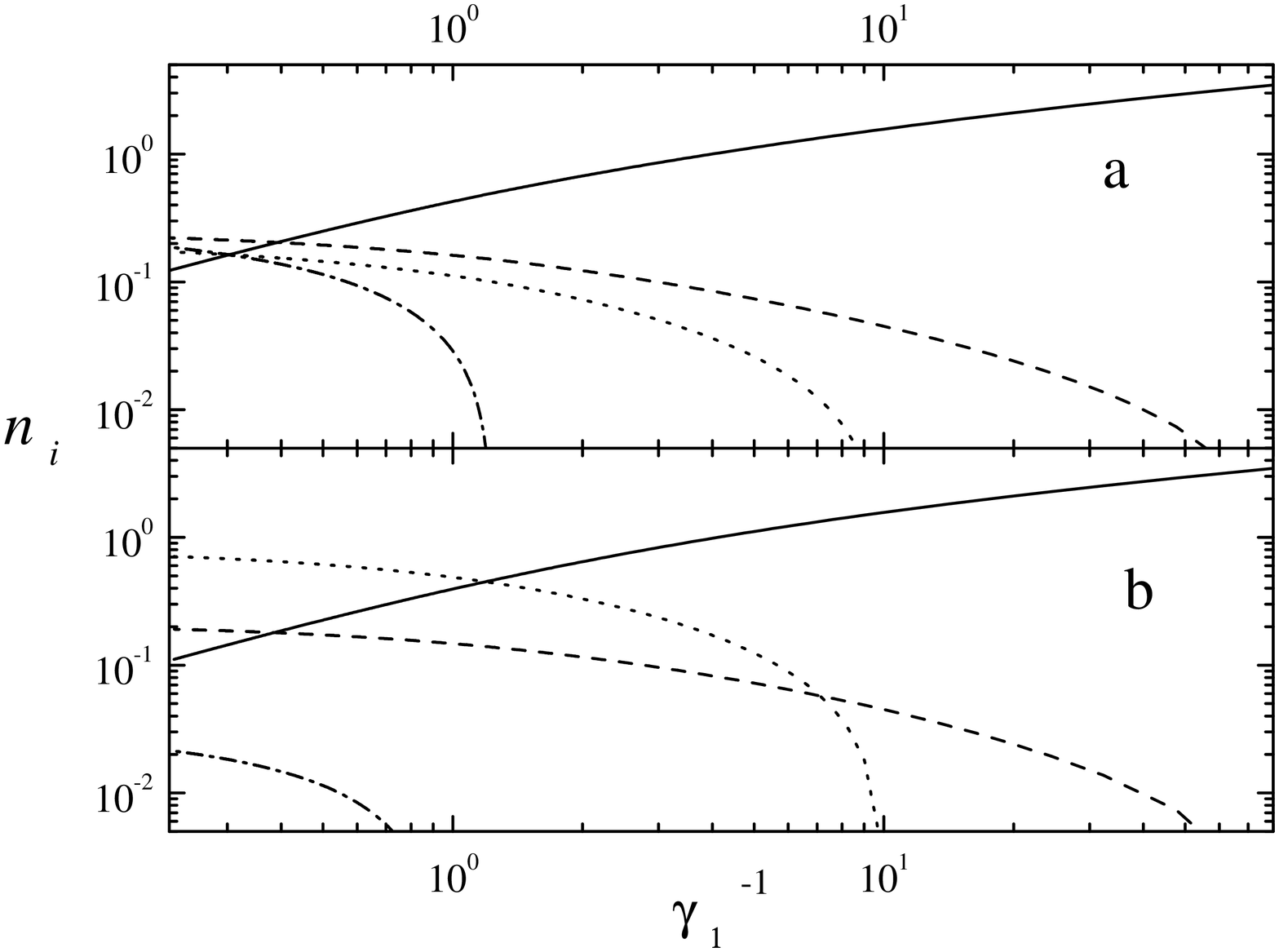}}
\caption{Asymptotic values of $n_i$ for different values of
$\gamma_1^{-1}$. In both cases  $\rho_1=5$, $\rho _2=2.5$, $\rho
_3=1.5$, $\rho _4=1$. Full line: Species 1; Dash line: Species 2;
Dotted line: Species 3; Dash-dotted line: Species 4. In a)
$\gamma_2=1$, $\gamma _3=0.5$ and $\gamma _4=0.1$,  in b) $\gamma
_2=1$, $\gamma _3=0.1$ and $\gamma _4=0.5$.} \label{f1}
\end{figure}

\section{Numerical Results}

As indicated before, here we focus in a case with $N$ small (in
fact $N=4$) that shows all the relevant aspects we can expect in
the large $N$ situation. Defining the ratio
$\rho_i=\frac{\beta_i}{\alpha _i}$ we consider several situations:
\\
a) when the different companies are in hierarchical order, that is
$\rho_1 >  \rho _2 > \rho _3 > \rho _4 $; \\
b) when we have $\rho_1 >  \rho _2 = \rho _3 = \rho _4$; \\
c) when $\rho_1 =  \rho _2 > \rho _3 = \rho _4$.

Among all the possible scenarios we have chosen those showing
regions of coexistence of all the species, thus in case (a) we
investigate the situation with fixed $\gamma _j$, with $j=2,3,4$
and varying $\gamma _1$. That allows us to have a control
parameter, but we recall that the choice is arbitrary. In this
case the usual strong and weak concepts make us consider the
species 1 as the strongest and the 4 as the weakest. With the
inclusion of the new term we find not only the possibility of
coexistence but also that the original hierarchical order can be
permuted several times as the parameter values are varied. Thus,
we find that the ranking of companies suffers many changes, with
companies interchanging roles several times. Some examples of this
case are shown in Fig.1. Besides these new features we observe the
classical extinction of companies as predicted by the exclusion
theorem. In all the cases the extinction occurs by one species at
a time. Figure 1 shows two typical results for the stationary
values reached by $n_j$ as function of $\gamma _1^{-1}$. It is
apparent that according to the different values, the relative
status of each company can change and even reversed with several
crossing among them. In this figure $\gamma _1$ varies
continuously from 10 to 0. The ratios $\rho_i=\frac{\beta
_i}{\alpha _i}$ are $\rho_1=5$, $\rho _2=2.5$, $\rho _3=1.5$,
$\rho _4=1$. In Fig.1a we have $\gamma _2=1$, $\gamma _3=0.5$ and
$\gamma _4=0.1$, while in Fig 1.b $\gamma _2=1$, $\gamma _3=0.1$
and $\gamma _4=0.5$.

\begin{figure}[h]
\centering \resizebox{\columnwidth}{!}{\includegraphics{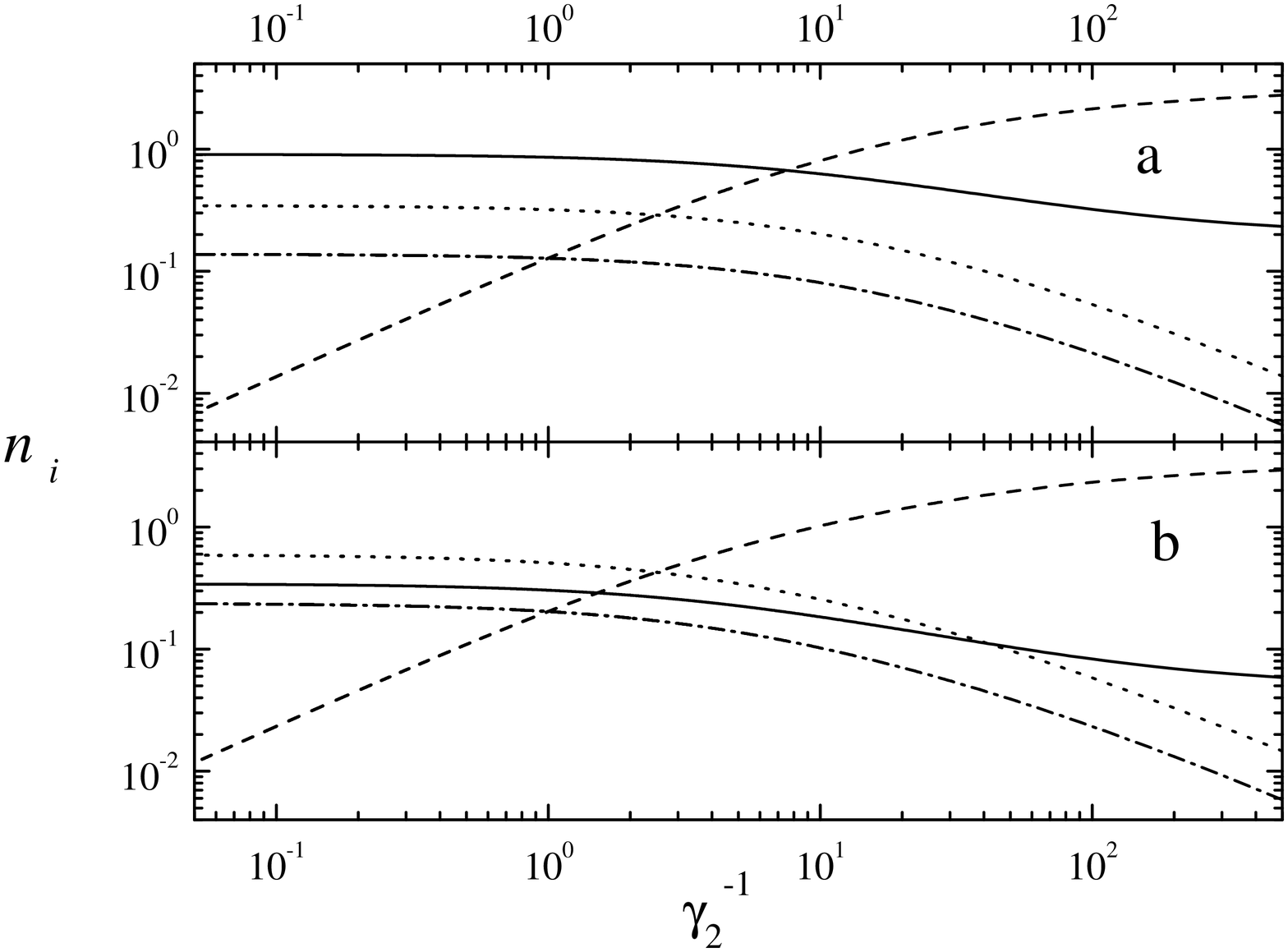}}
\caption{As in Figure 1 with  $\rho_1=4$, $\rho _2=\rho
_3=\rho_4=2$ and  a) $\gamma_1=.25$, $\gamma _3=0.5$ and $\gamma
_4=0.1$,  in b) $\gamma _1=1$, $\gamma _3=0.5$ and $\gamma
_4=0.1$.} \label{f2}
\end{figure}

For the case (b) above, we considered  as our variable $\gamma
_2$. In this case we have one strong species with $\rho_1=4$, and
three similar species, $\rho_2=\rho_3=\rho_4=2$, that could
coexist if there was not a strongest one. We want to note that
though $\rho_2=\rho_3=\rho_4$ the same is not true for $\alpha$
and $\beta$ values. Once again, we observe coexistence between
species and a reorganization of the company or agents ranking, not
accordingly to the original concept of strength but depending of
the values of $\gamma_i$. The coexistence is achieved within
certain parameter region. The extinction is gradual and governed
mainly by $\gamma$ values. The results are shown in Fig. 2, where
again we depict the stationary values of $n_j$ as functions of
$\gamma _x^{-1}$. In Fig 2.a we have $\gamma_1=0.25$,
$\gamma_3=0.5$, $\gamma_4=0.1$, while in Fig 2.b $\gamma_1=1.$,
$\gamma_3=0.5$, $\gamma_4=0.1$. It is apparent that the most
relevant parameter when considering competition is the value of
$\gamma_i$. When species are equally strong or weak, we observe
that different stationary density levels are reached according to
$\gamma_i$. If $\gamma_i$ is of the same order the coexistence is
granted. On the contrary, a species with a high $\gamma_i$ will
not survive even if competing with similar species. As an
additional feature we observe that a usual strong species $j$ will
not survive even if competing with weaker species if $\gamma_j$ is
much higher than that of the other species.

\begin{figure}[h]
\centering \resizebox{\columnwidth}{!}{\includegraphics{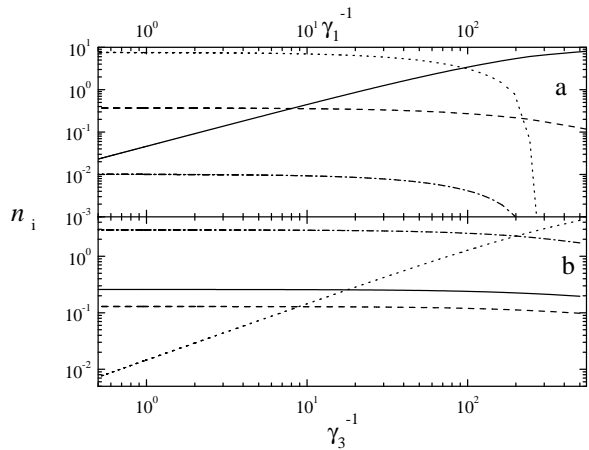}}
\caption{As in Figure 1 with  $\rho_1=\rho _2=4$, $\rho _3=\rho
_4=2$ and  a)$\gamma_2=0.25$, $\gamma _3=10^{-3}$ and $\gamma
_4=1.5$,  in b) $\gamma _1=0.5$, $\gamma _2=0.5$ and $\gamma
_4=0.01$} \label{f3}
\end{figure}

In case (c) we considered two situations: a first one  varying
$\gamma _1$; and a second varying $\gamma _3$. The results are
shown in Fig. 3, where again we depict the stationary values of
$n_j$ as functions of $\gamma _1^{-1}$ and $\gamma _3^{-1}$
respectively. In both cases 3 $\rho_1=\rho_2=4$,
$\rho_3=\rho_4=2$, while in a) $\gamma_2=0.25$, $\gamma_3=0.001$,
$\gamma_4=1.5$ and in b)$\gamma_1=0.5$, $\gamma_2=1.5$,
$\gamma_4=0.01$ . We confirm that the coexistence can be achieved
for proper $\gamma$ values. At the same time we observe that the
usual concept of strongest species cannot be applied as in both
previous cases, species 2, one of the strongest if $\gamma_i$ were
zero remains as the weakest species due to a high $\gamma_2$
value.

\section{Conclusions}

The results shown above indicates how the study of simplified
models could help in the understanding of the role played by
$\gamma $, an internal company parameter, associated to the
company efficiency, in situations where the complexity of the
economic reality makes very hard to obtain a complete model. In
our case, the model so far studied could give some hints on the
behavior of systems of companies in competition and the possibility
of coexistence and the way that one company can use to
``eliminate" the competing ones adopting a policy tending to
adequately change its $\gamma$. Here we have analyzed the effect
of explicitly including the {\it carrying capacity} of the
environment within our {\it toy} model for describing the
coexistence of species in competition. The results put in evidence
the role played by the term associated to $\gamma$ in the
possibility of coexistence. We recall that in the absence of such
term the exclusion principle is valid and only the species (one or
more) with the highest $\rho$ survive. It is also this term that,
within certain parameter region, governs the company or agents
ranking. A model written in the same terms can clearly also be
applicable to ecological situations. But in this case, rather than
a deterministic control of $\gamma$, some fluctuations or cyclic
changes in this parameter should be considered. This is the
subject of a work of us in progress. \\

{\bf Acknowledgements:}  Partial support from CONICET and ANPCyT,
both Argentina agencies, as well as to Fundaci\'on Antorchas, are
greatly acknowledged. HSW wants to thank to Iberdrola S.A.,
Spain, for an award within the {\it Iberdrola Visiting Professor
Program in Science and Technology}, and to the IMEDEA and
Universitat de les Illes Balears, Palma de Mallorca, Spain, for
the kind hospitality extended to him.

\end{document}